\documentclass[reprint, amsmath,amssymb, aps, prl, shortbibliography, lengthcheck,superscriptaddress]{revtex4-1}

\usepackage{graphicx}
\usepackage{dcolumn}
\usepackage{bm}
\usepackage{hyperref}
\usepackage{epsfig}

\begin{document}

\title{Light-assisted ion-neutral reactive processes in the cold regime: \\ radiative molecule formation vs. charge exchange }

\author{Felix H.J. Hall}
\affiliation{Department of Chemistry, University of Basel, Klingelbergstrasse 80, 4056 Basel, Switzerland}
\author{Mireille Aymar}
\affiliation{Laboratoire Aim\'{e} Cotton, CNRS, B\^{a}timent 505, Universit\'{e} Paris-Sud XI, 91405 Orsay Cedex, France}
\author{Nadia Bouloufa-Maafa}
\affiliation{Laboratoire Aim\'{e} Cotton, CNRS, B\^{a}timent 505, Universit\'{e} Paris-Sud XI, 91405 Orsay Cedex, France}
\affiliation{ Universit$\acute{e}$ Cergy-Pontoise, 95000 Cergy-Pontoise, France.}
\author{Olivier Dulieu}
\email{olivier.dulieu@lac.u-psud.fr}
\affiliation{Laboratoire Aim\'{e} Cotton, CNRS, B\^{a}timent 505, Universit\'{e} Paris-Sud XI, 91405 Orsay Cedex, France}
\author{Stefan Willitsch}
\email{stefan.willitsch@unibas.ch}
\affiliation{Department of Chemistry, University of Basel, Klingelbergstrasse 80, 4056 Basel, Switzerland}

\date{\today}

\begin{abstract}
We present a combined experimental and theoretical study of cold reactive collisions between laser-cooled Ca$^+$ ions and Rb atoms in an ion-atom hybrid trap. We observe rich chemical dynamics which are interpreted in terms of non-adiabatic and radiative charge exchange as well as radiative molecule formation using high-level electronic structure calculations. We study the role of light-assisted processes and show that the efficiency of the dominant chemical pathways is considerably enhanced in excited reaction channels. Our results illustrate the importance of radiative and non-radiative processes for the cold chemistry occurring in ion-atom hybrid traps.
\end{abstract}

\maketitle

Over the past few years, impressive progress has been achieved in the study of reactive collisions at ultralow energies $E_\text{coll}/k_\text{B} \ll 1$~K \cite{bell09b, dulieu09a}. Recent landmark studies using neutral molecules highlighted the distinct quantum character of reactive processes in this regime and demonstrated new approaches for an unprecedented control of molecular collisions \cite{ni10a, ospelkaus10b}. Ion-neutral reactions are another class of processes which exhibit different long-range interactions and therefore a different chemical behavior in comparison to neutrals \cite{cote00a, bodo02a, willitsch08a, willitsch08b, bell09a, idziaszek09a, gao10a}. With the development of hybrid traps in which laser-cooled atomic ions stored in a radiofrequency ion trap are combined with ultracold neutral atoms in a magneto-optical trap \cite{smith05a, grier09a, rellergert11a} or a Bose-Einstein-condensate \cite{zipkes10a, schmid10a}, the study of ion-neutral reactions in the energy range between 1 and 10$^{-3}$~Kelvin (usually termed the ``cold'' regime) has recently become possible. Under these conditions, only a few partial waves contribute to the collision so that resonance as well as radiative effects can become important \cite{weiner99a, cote00a,bodo02a,idziaszek09a}. 

One key question pertains to the types of chemical processes which can occur in hybrid traps. So far, either fast near-resonant homonuclear charge exchange (in Yb-Yb$^+$ \cite{grier09a}) or a slow loss of the atomic ions from the trap were observed (in Rb-Yb$^+$ \cite{zipkes10a} and Rb-Ba$^+$ \cite{schmid10a}). For Rb-Yb$^+$, the latter observation was rationalized in terms of radiative and non-radiative charge exchange \cite{zipkes10b}. The feasibility of molecular-ion formation has also been considered, and evidence for a radiative mechanism has recently been found in the Ca-Yb$^+$ system \cite{rellergert11a}. However, a general understanding of the interplay between these reactive processes and in particular the role of light remains to be established. 

In the current study, we present a combined experimental and theoretical study of ion-neutral reactive collisions in a Rb-Ca$^+$ hybrid trap. Our experimental results are interpreted using high-level electronic structure calculations of the CaRb$^+$ potential energy curves (PECs) up to the twenty-second dissociation limit. We observe rich chemical dynamics which we rationalize in terms of non-adiabatic and radiative effects. We show that the efficiency of the dominant chemical processes (radiative molecule formation, radiative and non-radiative charge exchange) is considerably enhanced in excited reaction channels populated in the presence of radiation. Using Rb-Ca$^+$ as a model system, our results illustrate the reactive processes which can occur under the ``cold'' conditions of ion-atom hybrid traps.

\begin{figure}[t]
\epsfig{file=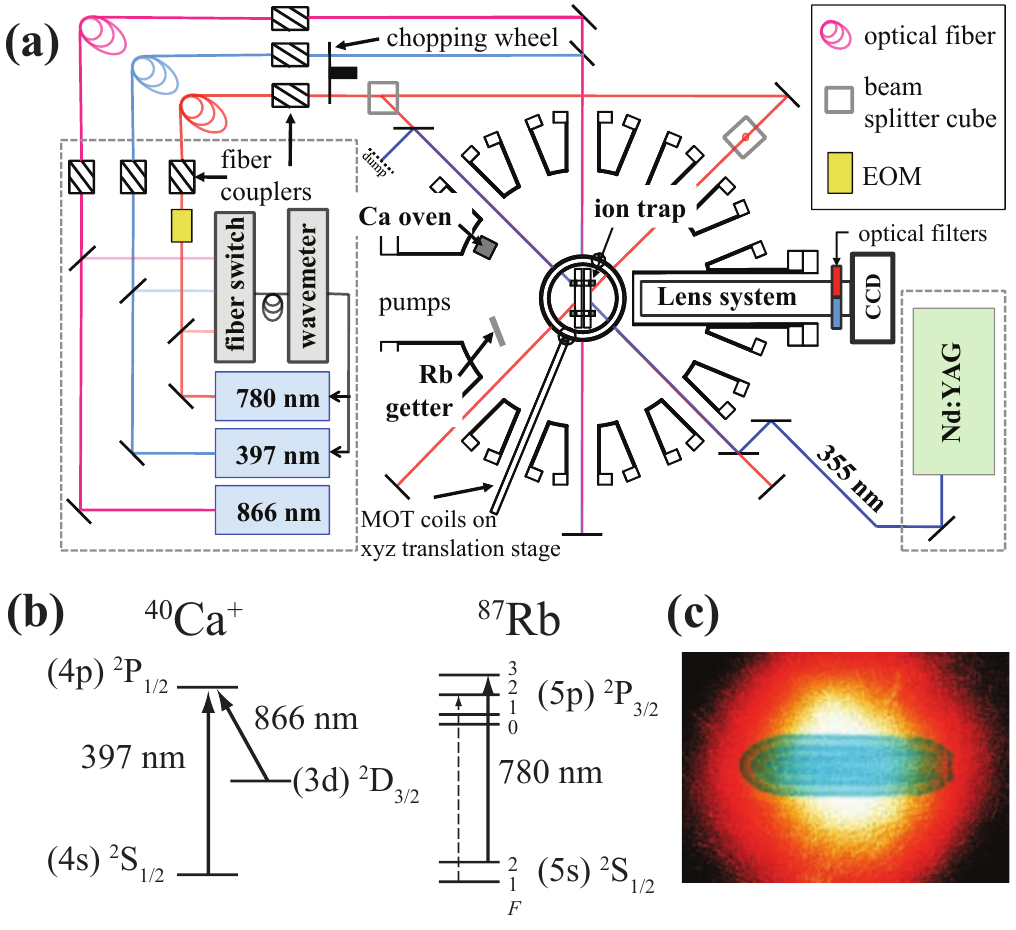,width=\columnwidth}
\caption{\label{fig1} (a) Schematic of the experimental setup. (b) Laser-cooling schemes for $^{40}$Ca$^+$ and $^{87}$Rb. (c) Superposed fluorescence images of a Ca$^+$ ion Coulomb crystal (blue) and a cloud of ultracold Rb atoms (red) in the hybrid trap.}
\end{figure}

For the present study, an ion-atom hybrid trap was implemented by superimposing a linear radiofrequency ion trap \cite{willitsch08b} for laser cooling $^{40}$Ca$^+$ ions with a magneto-optical trap (MOT) \cite{raab87a} for $^{87}$Rb atoms (Figure \ref{fig1} (a)). Doppler laser cooling of the Ca$^+$ ions was achieved using two diode laser beams at 397~nm and 866~nm pumping on the $(4s)~^2S_{1/2}\rightarrow(4p)~^2P_{1/2}$ and $(3d)~^2D_{3/2}\rightarrow(4p)~^2P_{1/2}$ transitions, respectively (Figure \ref{fig1} (b)). Upon laser cooling, the ions form Coulomb crystals \cite{willitsch08b}. The average ion kinetic energies were dominated by the micromotion (the fast motion driven by the radiofrequency trapping fields) whose contribution was characterized by a comparison of experimental Coulomb crystal images with molecular-dynamics simulations \cite{willitsch08b, bell09a}. In our experiments, the average collision energies $\langle E_\text{coll}\rangle$ were entirely governed by the ion kinetic energies which were varied by changing the size and shape of the Coulomb crystals as discussed in Ref. \cite{bell09a}.

The MOT was set up with two water-cooled solenoids installed in vacuum to generate a quadrupolar magnetic field with a gradient of 20 G/cm. Laser beams around 780~nm in an optical-molasses configuration were used for cooling and repumping Rb on the $(5s)~^2S_{1/2}\rightarrow(5p)~^2P_{3/2}$ transition (see Fig. \ref{fig1} (b)). The number density and temperature of the Rb atoms in the MOT were established using standard fluorescence measurement and time-of-flight methods, respectively \cite{pradhan08a}. The fluorescence of either the ions or the neutral atoms was isolated using narrow-bandpass color filters and imaged onto a CCD camera (Figure \ref{fig1} (c)). A detailed description of the experimental apparatus and procedures will be given in a subsequent publication \cite{hall11b}.

In our experiment, the Ca$^+$ and Rb cooling lasers were alternately blocked using a mechanical chopper at a frequency of 1000 Hz in order to prevent photoionization of Rb out of the $(5p)~^2P_{3/2}$ level by 397~nm photons. Under these conditions, the typical Rb number densities and temperatures achieved were on the order of $1\times10^{9}$~cm$^{-3}$ and $T=150-200~\mu$K, respectively. The chopping of the cooling-laser beams only negligibly affected the kinetic energies of the ions and their overlap with the Rb cloud.

\begin{figure}[t]
\epsfig{file=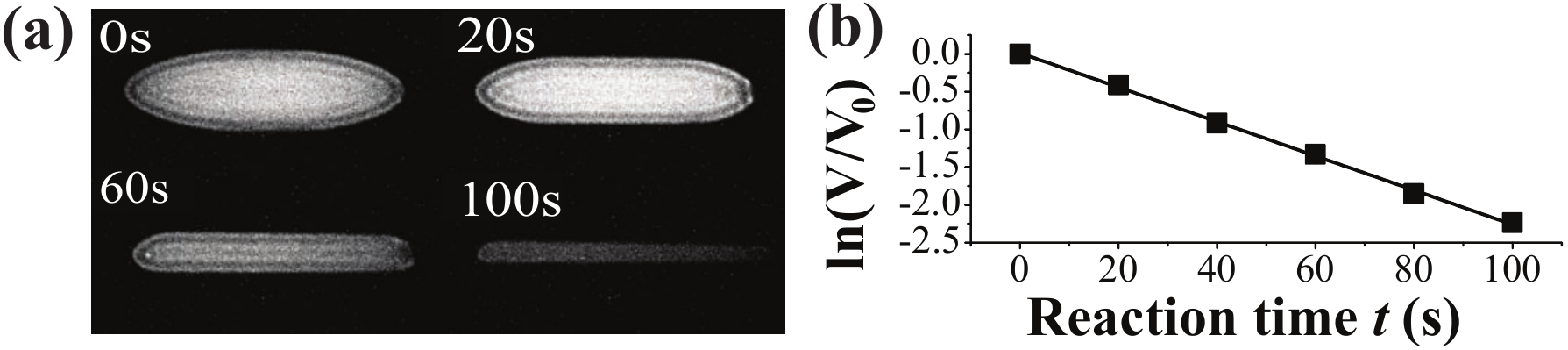,width=\columnwidth}
\caption{\label{fig2} (a) Series of Ca$^+$ ion laser-cooling fluorescence images as a function of the time of reaction with ultracold Rb atoms. (b) Pseudo-first-order analysis of the reaction kinetics of the experiment shown in (a). The uncertainties are smaller than the size of the symbols in the plot.}
\end{figure}

Reactive collisions between Ca$^+$ and Rb lead to a decrease in the number of Ca$^+$ ions in the Coulomb crystals as shown in Figure \ref{fig2} (a). When both species are laser cooled, the populations are distributed over the $(4s)~^2S_{1/2},~(4p)~^2P_{1/2}$ and $(3d)~^2D_{3/2}$ states of Ca$^+$ and the $(5s)~^2S_{1/2}$ and $(5p)~^2P_{3/2}$ states of Rb (compare Fig. \ref{fig1} (b)). Reactive collisions occur in excited states of the Rb-Ca$^+$ system and the observed reaction rates represent an average over all possible channels. Note that simultaneous excitation of both species does not occur because of the alternate chopping of the cooling laser beams.  

Rate coefficients were determined by measuring the decrease of the Ca$^+$ Coulomb crystal volume $V$ as a function of the reaction time and fitting the results to a pseudo-first-order rate expression $\ln(V/V_0)=k^\prime t$ \cite{willitsch08a} where $V_0$ denotes the initial crystal volume and $t$ is the reaction time, see Figure \ref{fig2} (b). Second-order rate coefficients $k$ were obtained by dividing the pseudo first-order rate constants $k^\prime$ by the average density of Rb atoms $N_\text{Rb}$ in the MOT: $k=k^\prime/N_\text{Rb}$. For the measurement displayed in Figure \ref{fig2}, the rate coefficient was established to be $k=2.5(9)\times10^{-11}~\mathrm{cm^3~s^{-1}}$, only two orders of magnitude smaller than the collisional (Langevin) rate coefficient ($k_\text{L}=3\times10^{-9}~\mathrm{cm^3~s^{-1}}$ \cite{tacconi11a}).

When the Ca$^+$ cooling lasers were switched off, a rate coefficient $k_s=3(1)\times10^{-12}$~cm$^3$~s$^{-1}$ was obtained. This value proved to be insensitive to the excited-state population of Rb which was varied in the range from 2\% to 5\% compatible with a stable operation of the MOT. However, the ions are likely to heat up in the absence of laser cooling which also reduces their overlap with the cold-atom cloud. Therefore, an accurate value for collision energies cannot be given under these conditions and this value for $k_s$ must be regarded as an estimate of the rate coefficient in the lowest reaction channel Ca$^+(4s)$+Rb$(5s)$.

\begin{figure}[t]
\epsfig{file=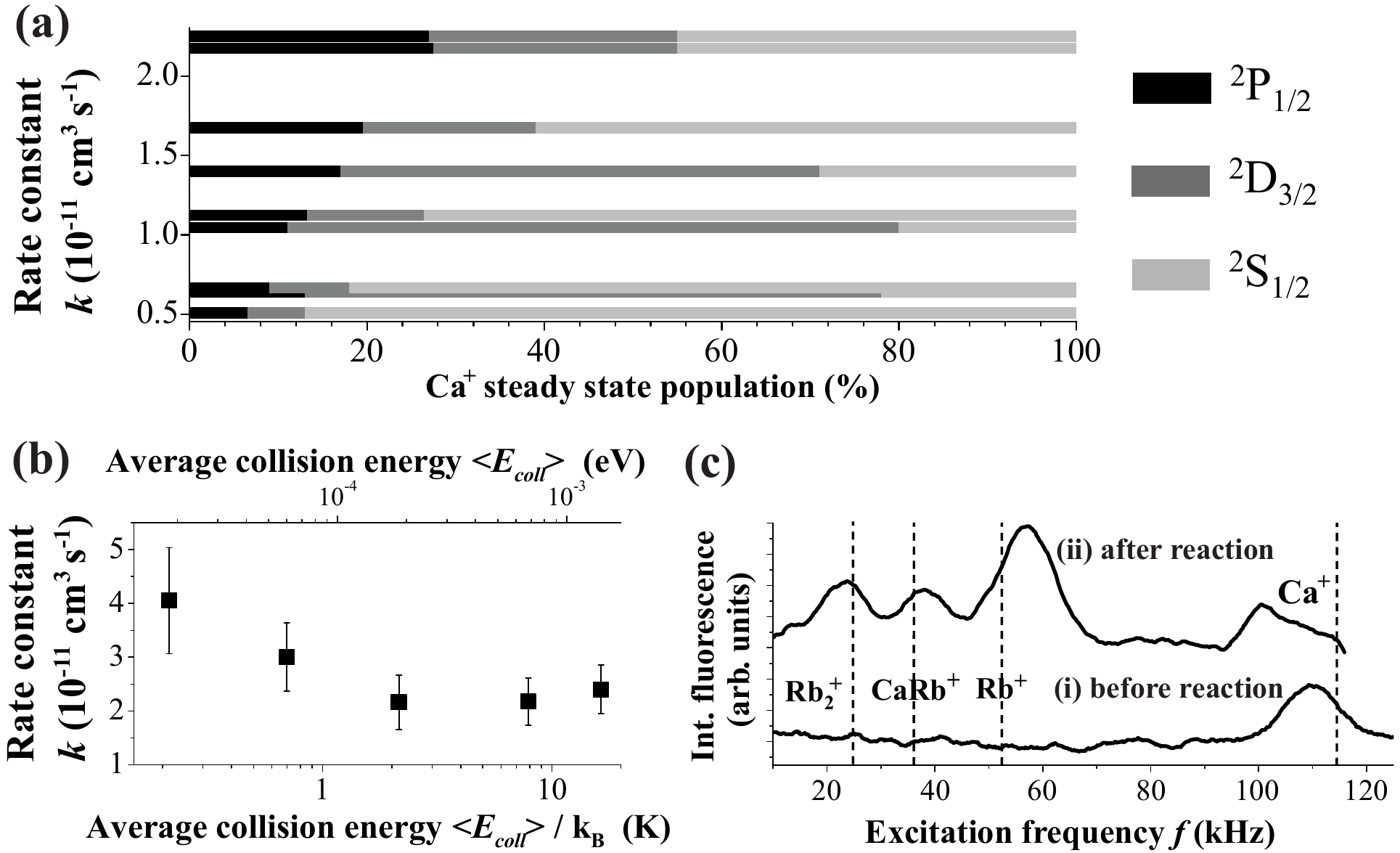,width=\columnwidth}
\caption{\label{fig3} (a) Variation of reaction rate constants $k$ with Ca$^+$ level populations. The bars indicate the relative level populations obtained by varying the detunings of the cooling-laser beams, their intercepts with the y axis give the values of the corresponding rate constants. Each data point corresponds to an average of three consecutive measurements (average statistical uncertainty (2$\sigma$) $\Delta k=3\times10^{-12}$~cm$^{3}$~s$^{-1}$). (b) Rate constant as a function of the average collision energy $\langle E_\text{coll}\rangle/k_\text{B}$. The error bars denote the statistical uncertainty ($2\sigma$) of the measurements. (c) Resonant-excitation mass spectra of Coulomb crystals (i) before reaction, (ii) after reaction. The dashed vertical lines indicate single-ion motional frequencies.}
\end{figure}

To assess the contribution of the excited Ca$^+$ channels, we modified the populations in the Ca$^+$ states by varying the frequency detuning of the 397~nm and 866~nm lasers in the range 20-100~MHz and 0-70~MHz, respectively. In these measurements, it was ensured that the ion cloud always remained Coulomb-crystallized so that the effect of the detuning on the average ion energies is negligible. Figure \ref{fig3} (a) shows the resulting rate coefficients which clearly correlate with the population in the Ca$^+(4p)~^2P_{1/2}$ state, whereas no correlation with the populations in the $(4s)~^2S_{1/2}$ and $(3d)~^2D_{3/2}$ levels is observed. The populations were inferred from an Einstein rate-equation model of the laser excitations in the Ca$^+$ three-level system, with an estimated relative uncertainty of 20\% \cite{hall11b}.

We fitted the data in Fig. \ref{fig3} (a) to a kinetic model accounting for the relevant reaction channels. Taking into account the chopping of the Ca$^+$ cooling laser beam, the measured rate coefficient can be expressed as $k=\tfrac{1}{2}\big[k_s(p_s+1)+k_pp_p+k_dp_d\big]$. Here, $k_s, k_p, k_d$ and $p_s, p_p,p_d$ stand for the rate coefficients and populations in the Ca$^+(4s), (4p)$ and $(3d)$ levels. We find $k_p=1.5(6)\times10^{-10}$~cm$^3$~s$^{-1}$ and $k_{s,d}\leq3\times10^{-12}$~cm$^3$~s$^{-1}$. 

Rate coefficients as a function of the collision energy are displayed in Fig. \ref{fig3} (b). Within the range  $\langle E_\text{coll}\rangle/k_\text{B}=$200~mK-20~K, we observe an increase by a factor of two. A detailed analysis of these observations will be published in a subsequent article \cite{hall11b}.

The chemical identity of the reaction products was established using resonant-excitation mass spectrometry of the Coulomb crystals \cite{roth09a}. The spectra taken before immersing the Ca$^+$ Coulomb crystal in the MOT only show a single resonance corresponding to the excitation of the Ca$^+$ ions (Figure \ref{fig3} (c) (i)). By contrast, the spectra recorded after the reaction (Figure \ref{fig3} (c) (ii)) exhibit in total four distinct resonances which are identified as Ca$^+$, Rb$^+$, CaRb$^+$ and Rb$_2^+$ by comparison with the expected single-ion excitation frequencies (indicated by dashed vertical lines). 

Whereas the Rb$^+$ product is a result of charge exchange between Ca$^+$ and Rb, CaRb$^+$ can only be formed in an association reaction whereby the collision complex is stabilized either collisionally or by the spontaneous emission of a photon (radiative association, RA). Collisional stabilization is negligible at the low Rb densities in the MOT. Moreover, Rb$_2^+$ is only observed in the presence of CaRb$^+$ indicating that this product results from the reaction CaRb$^+$+Rb$\rightarrow$ Rb$_2^+$ + Ca which is exothermic by $\approx$0.4~eV according to our calculations. Product-ion peaks for Rb$^+$ and CaRb$^+$ were observed over the range of cooling laser detunings used in the present study and when the Ca$^+$ cooling laser was switched off, indicating that these ions are produced in the ground as well as in the excited reaction channels \cite{hall11b}.

\begin{figure}[t]
\epsfig{file=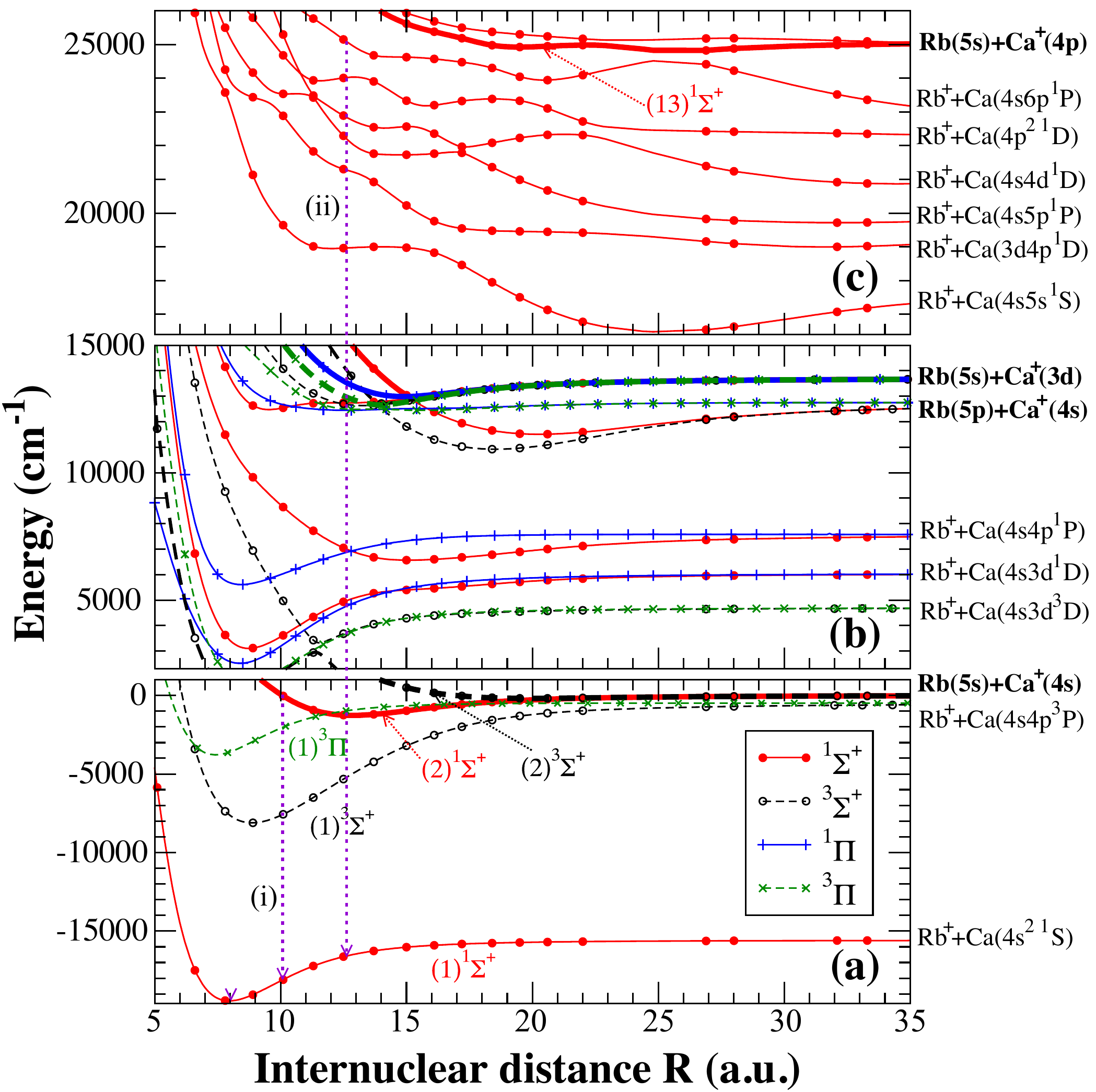,width=\columnwidth}
\caption{\label{fig4} Computed RbCa$^+$ potential energy curves (PECs) in the regions of (a) Rb($5s$)+Ca$^+$($4s$), (b) Rb($5s$)+Ca$^+$($3d$), (c) Rb($5s$)+Ca$^+$($4p$). The PECs of the relevant entrance channels of the reaction are highlighted with thick lines. PECs for $^1\Delta$ and $^3\Delta$ symmetries are omitted for clarity. Downward arrows suggest possible pathways for formation of ground-state RbCa$^+$ ions by radiative association. In (c), only PECs of $^1\Sigma^+$ symmetry are displayed for clarity.}
\end{figure}

Figure \ref{fig4} displays the RbCa$^+$ potential energy curves (without spin-orbit interaction) up to the twenty-second dissociation threshold Rb($5s$)+Ca$^+$($4p$) which can be reached in the present light-induced dynamics. Currently, the only reliable way to describe such highly-excited states consists in treating explicitly only the two valence electrons of CaRb$^+$ in the field of two ionic cores Rb$^+$ and Ca$^{2+}$ represented by effective core potentials \cite{guerout10a, guerout11a}. Correlation between core and valence electrons is modeled via effective core polarization potentials. Therefore, a full configuration interaction (FCI) calculation is achievable in a configuration space built from a large basis set of Gaussian orbitals for the valence electrons \cite{guerout11a, bouissou10a}. The complete set of calculations will be presented elsewhere \cite{hall11b}.

Over the energy range of about 41000~cm$^{-1}$, two kinds of asymptotic channels are present: a Rb ($5s$ or $5p$) atom colliding with a Ca$^+$ ($4s$, $3d$, or $4p$) ion (4 channels), and a Rb$^+$ ion colliding with a Ca atom (18 channels). In our approach, the level energies of Rb and Ca$^+$ are constrained to the experimental values. The energies of the effective two-electron Ca atom are obtained through the FCI calculation, with an accuracy of a few hundred wave numbers typical for such calculations. Therefore, in the congested region around the Rb($5s$)+Ca$^+$($4p$) asymptote (Figure \ref{fig4} (c)) the order of several asymptotic limits is changed, preventing a detailed state-to-state modeling of the dynamics. Nevertheless, the calculations provide a useful guide to a qualitative understanding of the underlying reaction mechanisms.

Figure \ref{fig4} (a) shows the region already studied theoretically in Ref. \cite{tacconi11a}. Note that the Rb($5s$)+Ca$^+$($4s$) entrance channel is not the lowest dissociation limit. Non-radiative charge exchange (NRCE) induced by non-adiabatic couplings can occur around the crossings with the $(1)^3\Pi$ state correlating with the Rb$^+$+Ca$(4s4p~^3P)$ asymptote \cite{tacconi11a}. Moreover, radiative relaxation from the $(2)^1\Sigma^+$ entrance channel to the $(1)^1\Sigma^+$ state can either lead to radiative charge exchange (RCE) or RA forming molecular ions provided that the relevant Franck-Condon (FC) factors are sufficiently large. The dashed arrow (i) in Figure \ref{fig4} (a) suggests a transition for which favorable FC overlap for RA can be expected. This picture is compatible with our experimental observation of both RA and CE products in this channel. We note that the upper limit for the experimental rate coefficient $k_s\leq 3\times10^{-12}$~cm$^3$~s$^{-1}$ determined above is close to value of $\approx 1\times10^{-12}$~cm$^3$~s$^{-1}$ estimated from theoretical NRCE cross sections reported in Ref. \cite{tacconi11a}. 

In the case of the Rb($5s$)+Ca$^+$($3d$) entrance channel, no close-lying molecular state asymptotically correlating with neutral Ca exist which would enable efficient NRCE (Figure \ref{fig4} (b)). This is in agreement with our experimental finding of a small reaction rate for this channel ($k_{d}\leq3\times10^{-12}$~cm$^3$~s$^{-1}$). RCE and RA connecting to lower-lying states are in principle possible in various symmetries. However, inspection of the relevant PECs suggests that the FC factors are small. 

Figure \ref{fig4} (c) shows the complicated network of the PECs for the $^1\Sigma^+$ states in the region around the Rb$(5s)$+Ca$^+(4p)$ entrance channel resulting from the large number of dissociation limits within a small energy range. The situation is comparable for the other symmetries which are not shown for clarity. From Figure \ref{fig4} (c), a mechanism explaining the large reaction rates in this channel becomes apparent. At large distances, the Rb($5s$)+Ca$^+$($4p$) PEC is attractive (varying as $-R^{-4}$) and isolated from other channels with the same asymptotic charge state, e.g., Rb($5s$)+Ca$^+$($4s, 3d$). By contrast, while the long-range PECs of the states correlating with the Rb$^+$+Ca$^*$ asymptotes are also attractive, they soon turn into repulsive curves as being Rydberg states converging to the Rb$^+$+Ca$^+$ ionization threshold. As a consequence, the PEC of the entrance channel (the (13)$^1\Sigma^+$ state) undergoes avoided crossings with lower curves correlating with charge-transfer asymptotes. Non-adiabatic transitions at these crossings will result in a large rate for NRCE leading to the formation of Rb$^+$. Moreover, through consecutive non-adiabatic transitions in the intricate network of avoided crossings visible in Fig. \ref{fig4} (c), the collision partners can approach to short internuclear distances where many channels are open for RCE and RA in all symmetries. A possible radiative relaxation pathway to the absolute ground state is suggested by the downward arrow (ii). Additionally, radiative relaxation at long range can also contribute to the observed charge-exchange and molecule-formation processes.

This mechanism for radiative association and charge transfer from excited states is reminiscent of the processes leading to ``radiative escape'' of ultracold atoms from MOTs \cite{weiner99a}: excitation occurs in the asymptotic region in the early stages of the collision, followed by radiative relaxation at shorter range. In general, the multitude of different charge-transfer channels in ion-atom hybrid systems leads to a considerable increase of the density of states in comparison to purely neutral collision systems. This situation implies a higher probability for favorable Franck-Condon overlaps which increases the possibilities (and therefore the rates) for RA and RCE out of excited channels. This situation can be generalized to all alkaline atom-alkaline earth ion systems and it can thus be expected that RA and RCE are generally important processes in the collisional dynamics of these species. 

In summary, our results illustrate the rich chemical dynamics exhibited by even the simplest possible (i.e., atomic) ion-neutral collision systems in the cold regime. In particular, we highlighted the complex interplay between non-adiabatic effects, radiative charge exchange and molecule formation. Our findings underline the importance of light in enhancing these processes and suggest that efficient photoassociation of molecular ions using red detuned laser radiation might be feasible in the Rb-Ca$^+$ system.

This work was supported by the Swiss National Science Foundation (grant nr. PP0022\_118921) and the University of Basel. Laboratoire Aim\'e Cotton is a member of the Institut Francilien de Recherches sur les Atomes Froids (IFRAF). We thank Prof. Francesco Gianturco for communicating results prior to publication.

\bibliographystyle{apsrev4-1}
\bibliography{../../../../docs/bib_file/main}

\begin{thebibliography}{26}%
\makeatletter
\providecommand \@ifxundefined [1]{%
 \@ifx{#1\undefined}
}%
\providecommand \@ifnum [1]{%
 \ifnum #1\expandafter \@firstoftwo
 \else \expandafter \@secondoftwo
 \fi
}%
\providecommand \@ifx [1]{%
 \ifx #1\expandafter \@firstoftwo
 \else \expandafter \@secondoftwo
 \fi
}%
\providecommand \natexlab [1]{#1}%
\providecommand \enquote  [1]{``#1''}%
\providecommand \bibnamefont  [1]{#1}%
\providecommand \bibfnamefont [1]{#1}%
\providecommand \citenamefont [1]{#1}%
\providecommand \href@noop [0]{\@secondoftwo}%
\providecommand \href [0]{\begingroup \@sanitize@url \@href}%
\providecommand \@href[1]{\@@startlink{#1}\@@href}%
\providecommand \@@href[1]{\endgroup#1\@@endlink}%
\providecommand \@sanitize@url [0]{\catcode `\\12\catcode `\$12\catcode
  `\&12\catcode `\#12\catcode `\^12\catcode `\_12\catcode `\%12\relax}%
\providecommand \@@startlink[1]{}%
\providecommand \@@endlink[0]{}%
\providecommand \url  [0]{\begingroup\@sanitize@url \@url }%
\providecommand \@url [1]{\endgroup\@href {#1}{\urlprefix }}%
\providecommand \urlprefix  [0]{URL }%
\providecommand \Eprint [0]{\href }%
\providecommand \doibase [0]{http://dx.doi.org/}%
\providecommand \selectlanguage [0]{\@gobble}%
\providecommand \bibinfo  [0]{\@secondoftwo}%
\providecommand \bibfield  [0]{\@secondoftwo}%
\providecommand \translation [1]{[#1]}%
\providecommand \BibitemOpen [0]{}%
\providecommand \bibitemStop [0]{}%
\providecommand \bibitemNoStop [0]{.\EOS\space}%
\providecommand \EOS [0]{\spacefactor3000\relax}%
\providecommand \BibitemShut  [1]{\csname bibitem#1\endcsname}%
\let\auto@bib@innerbib\@empty
\bibitem [{\citenamefont {Bell}\ and\ \citenamefont {Softley}(2009)}]{bell09b}%
  \BibitemOpen
  \bibfield  {author} {\bibinfo {author} {\bibfnamefont {M.~T.}\ \bibnamefont
  {Bell}}\ and\ \bibinfo {author} {\bibfnamefont {T.~P.}\ \bibnamefont
  {Softley}},\ }\href@noop {} {\bibfield  {journal} {\bibinfo  {journal} {Mol.
  Phys.}\ }\textbf {\bibinfo {volume} {107}},\ \bibinfo {pages} {99} (\bibinfo
  {year} {2009})}\BibitemShut {NoStop}%
\bibitem [{\citenamefont {Dulieu}\ and\ \citenamefont
  {Gabbanini}(2009)}]{dulieu09a}%
  \BibitemOpen
  \bibfield  {author} {\bibinfo {author} {\bibfnamefont {O.}~\bibnamefont
  {Dulieu}}\ and\ \bibinfo {author} {\bibfnamefont {C.}~\bibnamefont
  {Gabbanini}},\ }\href@noop {} {\bibfield  {journal} {\bibinfo  {journal}
  {Rep. Prog. Phys.}\ }\textbf {\bibinfo {volume} {72}},\ \bibinfo {pages}
  {086401} (\bibinfo {year} {2009})}\BibitemShut {NoStop}%
\bibitem [{\citenamefont {Ni}\ \emph {et~al.}(2010)\citenamefont {Ni},
  \citenamefont {Ospelkaus}, \citenamefont {Wang}, \citenamefont
  {Qu\'{e}m\'{e}ner}, \citenamefont {Neyenhuis}, \citenamefont {\mbox{de
  Miranda}}, \citenamefont {Bohn}, \citenamefont {Ye},\ and\ \citenamefont
  {Jin}}]{ni10a}%
  \BibitemOpen
  \bibfield  {author} {\bibinfo {author} {\bibfnamefont {K.-K.}\ \bibnamefont
  {Ni}}, \bibinfo {author} {\bibfnamefont {S.}~\bibnamefont {Ospelkaus}},
  \bibinfo {author} {\bibfnamefont {D.}~\bibnamefont {Wang}}, \bibinfo {author}
  {\bibfnamefont {G.}~\bibnamefont {Qu\'{e}m\'{e}ner}}, \bibinfo {author}
  {\bibfnamefont {B.}~\bibnamefont {Neyenhuis}}, \bibinfo {author}
  {\bibfnamefont {M.~H.~G.}\ \bibnamefont {\mbox{de Miranda}}}, \bibinfo
  {author} {\bibfnamefont {J.~L.}\ \bibnamefont {Bohn}}, \bibinfo {author}
  {\bibfnamefont {J.}~\bibnamefont {Ye}}, \ and\ \bibinfo {author}
  {\bibfnamefont {D.~S.}\ \bibnamefont {Jin}},\ }\href@noop {} {\bibfield
  {journal} {\bibinfo  {journal} {Nature}\ }\textbf {\bibinfo {volume} {464}},\
  \bibinfo {pages} {1324} (\bibinfo {year} {2010})}\BibitemShut {NoStop}%
\bibitem [{\citenamefont {Ospelkaus}\ \emph {et~al.}(2010)\citenamefont
  {Ospelkaus}, \citenamefont {Ni}, \citenamefont {Wang}, \citenamefont
  {\mbox{de Miranda}}, \citenamefont {Neyenhuis}, \citenamefont
  {Qu{\'e}m{\'e}ner}, \citenamefont {Julienne}, \citenamefont {Bohn},
  \citenamefont {Jin},\ and\ \citenamefont {Ye}}]{ospelkaus10b}%
  \BibitemOpen
  \bibfield  {author} {\bibinfo {author} {\bibfnamefont {S.}~\bibnamefont
  {Ospelkaus}}, \bibinfo {author} {\bibfnamefont {K.-K.}\ \bibnamefont {Ni}},
  \bibinfo {author} {\bibfnamefont {D.}~\bibnamefont {Wang}}, \bibinfo {author}
  {\bibfnamefont {M.~H.~G.}\ \bibnamefont {\mbox{de Miranda}}}, \bibinfo
  {author} {\bibfnamefont {B.}~\bibnamefont {Neyenhuis}}, \bibinfo {author}
  {\bibfnamefont {G.}~\bibnamefont {Qu{\'e}m{\'e}ner}}, \bibinfo {author}
  {\bibfnamefont {P.~S.}\ \bibnamefont {Julienne}}, \bibinfo {author}
  {\bibfnamefont {J.~L.}\ \bibnamefont {Bohn}}, \bibinfo {author}
  {\bibfnamefont {D.~S.}\ \bibnamefont {Jin}}, \ and\ \bibinfo {author}
  {\bibfnamefont {J.}~\bibnamefont {Ye}},\ }\href@noop {} {\bibfield  {journal}
  {\bibinfo  {journal} {Science}\ }\textbf {\bibinfo {volume} {327}},\ \bibinfo
  {pages} {853} (\bibinfo {year} {2010})}\BibitemShut {NoStop}%
\bibitem [{\citenamefont {C\^{o}t\'{e}}\ and\ \citenamefont
  {Dalgarno}(2000)}]{cote00a}%
  \BibitemOpen
  \bibfield  {author} {\bibinfo {author} {\bibfnamefont {R.}~\bibnamefont
  {C\^{o}t\'{e}}}\ and\ \bibinfo {author} {\bibfnamefont {A.}~\bibnamefont
  {Dalgarno}},\ }\href@noop {} {\bibfield  {journal} {\bibinfo  {journal}
  {Phys. Rev. A}\ }\textbf {\bibinfo {volume} {62}},\ \bibinfo {pages} {012709}
  (\bibinfo {year} {2000})}\BibitemShut {NoStop}%
\bibitem [{\citenamefont {Bodo}\ \emph {et~al.}(2002)\citenamefont {Bodo},
  \citenamefont {Scifoni}, \citenamefont {Sebastianelli}, \citenamefont
  {Gianturco},\ and\ \citenamefont {Dalgarno}}]{bodo02a}%
  \BibitemOpen
  \bibfield  {author} {\bibinfo {author} {\bibfnamefont {E.}~\bibnamefont
  {Bodo}}, \bibinfo {author} {\bibfnamefont {E.}~\bibnamefont {Scifoni}},
  \bibinfo {author} {\bibfnamefont {F.}~\bibnamefont {Sebastianelli}}, \bibinfo
  {author} {\bibfnamefont {F.~A.}\ \bibnamefont {Gianturco}}, \ and\ \bibinfo
  {author} {\bibfnamefont {A.}~\bibnamefont {Dalgarno}},\ }\href@noop {}
  {\bibfield  {journal} {\bibinfo  {journal} {Phys. Rev. Lett.}\ }\textbf
  {\bibinfo {volume} {89}},\ \bibinfo {pages} {283201} (\bibinfo {year}
  {2002})}\BibitemShut {NoStop}%
\bibitem [{\citenamefont {Willitsch}\ \emph
  {et~al.}(2008{\natexlab{a}})\citenamefont {Willitsch}, \citenamefont {Bell},
  \citenamefont {Gingell}, \citenamefont {Procter},\ and\ \citenamefont
  {Softley}}]{willitsch08a}%
  \BibitemOpen
  \bibfield  {author} {\bibinfo {author} {\bibfnamefont {S.}~\bibnamefont
  {Willitsch}}, \bibinfo {author} {\bibfnamefont {M.~T.}\ \bibnamefont {Bell}},
  \bibinfo {author} {\bibfnamefont {A.~D.}\ \bibnamefont {Gingell}}, \bibinfo
  {author} {\bibfnamefont {S.~R.}\ \bibnamefont {Procter}}, \ and\ \bibinfo
  {author} {\bibfnamefont {T.~P.}\ \bibnamefont {Softley}},\ }\href@noop {}
  {\bibfield  {journal} {\bibinfo  {journal} {Phys. Rev. Lett.}\ }\textbf
  {\bibinfo {volume} {100}},\ \bibinfo {pages} {043203} (\bibinfo {year}
  {2008}{\natexlab{a}})}\BibitemShut {NoStop}%
\bibitem [{\citenamefont {Willitsch}\ \emph
  {et~al.}(2008{\natexlab{b}})\citenamefont {Willitsch}, \citenamefont {Bell},
  \citenamefont {Gingell},\ and\ \citenamefont {Softley}}]{willitsch08b}%
  \BibitemOpen
  \bibfield  {author} {\bibinfo {author} {\bibfnamefont {S.}~\bibnamefont
  {Willitsch}}, \bibinfo {author} {\bibfnamefont {M.~T.}\ \bibnamefont {Bell}},
  \bibinfo {author} {\bibfnamefont {A.~D.}\ \bibnamefont {Gingell}}, \ and\
  \bibinfo {author} {\bibfnamefont {T.~P.}\ \bibnamefont {Softley}},\
  }\href@noop {} {\bibfield  {journal} {\bibinfo  {journal} {Phys. Chem. Chem.
  Phys.}\ }\textbf {\bibinfo {volume} {10}},\ \bibinfo {pages} {7200} (\bibinfo
  {year} {2008}{\natexlab{b}})}\BibitemShut {NoStop}%
\bibitem [{\citenamefont {Bell}\ \emph {et~al.}(2009)\citenamefont {Bell},
  \citenamefont {Gingell}, \citenamefont {Oldham}, \citenamefont {Softley},\
  and\ \citenamefont {Willitsch}}]{bell09a}%
  \BibitemOpen
  \bibfield  {author} {\bibinfo {author} {\bibfnamefont {M.~T.}\ \bibnamefont
  {Bell}}, \bibinfo {author} {\bibfnamefont {A.~D.}\ \bibnamefont {Gingell}},
  \bibinfo {author} {\bibfnamefont {J.}~\bibnamefont {Oldham}}, \bibinfo
  {author} {\bibfnamefont {T.~P.}\ \bibnamefont {Softley}}, \ and\ \bibinfo
  {author} {\bibfnamefont {S.}~\bibnamefont {Willitsch}},\ }\href@noop {}
  {\bibfield  {journal} {\bibinfo  {journal} {Faraday Discuss.}\ }\textbf
  {\bibinfo {volume} {142}},\ \bibinfo {pages} {73} (\bibinfo {year}
  {2009})}\BibitemShut {NoStop}%
\bibitem [{\citenamefont {Idziaszek}\ \emph {et~al.}(2009)\citenamefont
  {Idziaszek}, \citenamefont {Calarco}, \citenamefont {Julienne},\ and\
  \citenamefont {Simoni}}]{idziaszek09a}%
  \BibitemOpen
  \bibfield  {author} {\bibinfo {author} {\bibfnamefont {Z.}~\bibnamefont
  {Idziaszek}}, \bibinfo {author} {\bibfnamefont {T.}~\bibnamefont {Calarco}},
  \bibinfo {author} {\bibfnamefont {P.~S.}\ \bibnamefont {Julienne}}, \ and\
  \bibinfo {author} {\bibfnamefont {A.}~\bibnamefont {Simoni}},\ }\href@noop {}
  {\bibfield  {journal} {\bibinfo  {journal} {Phys. Rev. A}\ }\textbf {\bibinfo
  {volume} {79}},\ \bibinfo {pages} {010702} (\bibinfo {year}
  {2009})}\BibitemShut {NoStop}%
\bibitem [{\citenamefont {Gao}(2010)}]{gao10a}%
  \BibitemOpen
  \bibfield  {author} {\bibinfo {author} {\bibfnamefont {B.}~\bibnamefont
  {Gao}},\ }\href@noop {} {\bibfield  {journal} {\bibinfo  {journal} {Phys.
  Rev. Lett.}\ }\textbf {\bibinfo {volume} {104}},\ \bibinfo {pages} {213201}
  (\bibinfo {year} {2010})}\BibitemShut {NoStop}%
\bibitem [{\citenamefont {Smith}\ \emph {et~al.}(2005)\citenamefont {Smith},
  \citenamefont {Makarov},\ and\ \citenamefont {Lin}}]{smith05a}%
  \BibitemOpen
  \bibfield  {author} {\bibinfo {author} {\bibfnamefont {W.~W.}\ \bibnamefont
  {Smith}}, \bibinfo {author} {\bibfnamefont {O.~P.}\ \bibnamefont {Makarov}},
  \ and\ \bibinfo {author} {\bibfnamefont {J.}~\bibnamefont {Lin}},\
  }\href@noop {} {\bibfield  {journal} {\bibinfo  {journal} {J. Mod. Opt.}\
  }\textbf {\bibinfo {volume} {52}},\ \bibinfo {pages} {2253} (\bibinfo {year}
  {2005})}\BibitemShut {NoStop}%
\bibitem [{\citenamefont {Grier}\ \emph {et~al.}(2009)\citenamefont {Grier},
  \citenamefont {Cetina}, \citenamefont {Oru\v{c}evi\'{c}},\ and\ \citenamefont
  {Vuleti\'{c}}}]{grier09a}%
  \BibitemOpen
  \bibfield  {author} {\bibinfo {author} {\bibfnamefont {A.~T.}\ \bibnamefont
  {Grier}}, \bibinfo {author} {\bibfnamefont {M.}~\bibnamefont {Cetina}},
  \bibinfo {author} {\bibfnamefont {F.}~\bibnamefont {Oru\v{c}evi\'{c}}}, \
  and\ \bibinfo {author} {\bibfnamefont {V.}~\bibnamefont {Vuleti\'{c}}},\
  }\href@noop {} {\bibfield  {journal} {\bibinfo  {journal} {Phys. Rev. Lett.}\
  }\textbf {\bibinfo {volume} {102}},\ \bibinfo {pages} {223201} (\bibinfo
  {year} {2009})}\BibitemShut {NoStop}%
\bibitem [{\citenamefont {Rellergert}\ \emph {et~al.}()\citenamefont
  {Rellergert}, \citenamefont {Sullivan}, \citenamefont {Kotochigova},
  \citenamefont {Petrov}, \citenamefont {Chen}, \citenamefont {Schowalter},\
  and\ \citenamefont {Hudson}}]{rellergert11a}%
  \BibitemOpen
  \bibfield  {author} {\bibinfo {author} {\bibfnamefont {W.~G.}\ \bibnamefont
  {Rellergert}}, \bibinfo {author} {\bibfnamefont {S.~T.}\ \bibnamefont
  {Sullivan}}, \bibinfo {author} {\bibfnamefont {S.}~\bibnamefont
  {Kotochigova}}, \bibinfo {author} {\bibfnamefont {A.}~\bibnamefont {Petrov}},
  \bibinfo {author} {\bibfnamefont {K.}~\bibnamefont {Chen}}, \bibinfo {author}
  {\bibfnamefont {S.~J.}\ \bibnamefont {Schowalter}}, \ and\ \bibinfo {author}
  {\bibfnamefont {E.~R.}\ \bibnamefont {Hudson}},\ }\href@noop {} {\bibinfo
  {journal} {arXiv:1104.5478 (preprint)}\ }\BibitemShut {NoStop}%
\bibitem [{\citenamefont {Zipkes}\ \emph
  {et~al.}(2010{\natexlab{a}})\citenamefont {Zipkes}, \citenamefont {Palzer},
  \citenamefont {Sias},\ and\ \citenamefont {K\"{o}hl}}]{zipkes10a}%
  \BibitemOpen
\bibfield  {journal} {  }\bibfield  {author} {\bibinfo {author} {\bibfnamefont
  {C.}~\bibnamefont {Zipkes}}, \bibinfo {author} {\bibfnamefont
  {S.}~\bibnamefont {Palzer}}, \bibinfo {author} {\bibfnamefont
  {C.}~\bibnamefont {Sias}}, \ and\ \bibinfo {author} {\bibfnamefont
  {M.}~\bibnamefont {K\"{o}hl}},\ }\href@noop {} {\bibfield  {journal}
  {\bibinfo  {journal} {Nature}\ }\textbf {\bibinfo {volume} {464}},\ \bibinfo
  {pages} {388} (\bibinfo {year} {2010}{\natexlab{a}})}\BibitemShut {NoStop}%
\bibitem [{\citenamefont {Schmid}\ \emph {et~al.}(2010)\citenamefont {Schmid},
  \citenamefont {H\"{a}rter},\ and\ \citenamefont {\mbox{Hecker
  Denschlag}}}]{schmid10a}%
  \BibitemOpen
  \bibfield  {author} {\bibinfo {author} {\bibfnamefont {S.}~\bibnamefont
  {Schmid}}, \bibinfo {author} {\bibfnamefont {A.}~\bibnamefont {H\"{a}rter}},
  \ and\ \bibinfo {author} {\bibfnamefont {J.}~\bibnamefont {\mbox{Hecker
  Denschlag}}},\ }\href@noop {} {\bibfield  {journal} {\bibinfo  {journal}
  {Phys. Rev. Lett.}\ }\textbf {\bibinfo {volume} {105}},\ \bibinfo {pages}
  {133202} (\bibinfo {year} {2010})}\BibitemShut {NoStop}%
\bibitem [{\citenamefont {Weiner}\ \emph {et~al.}(1999)\citenamefont {Weiner},
  \citenamefont {Bagnato}, \citenamefont {Zilio},\ and\ \citenamefont
  {Julienne}}]{weiner99a}%
  \BibitemOpen
  \bibfield  {author} {\bibinfo {author} {\bibfnamefont {J.}~\bibnamefont
  {Weiner}}, \bibinfo {author} {\bibfnamefont {V.~S.}\ \bibnamefont {Bagnato}},
  \bibinfo {author} {\bibfnamefont {S.}~\bibnamefont {Zilio}}, \ and\ \bibinfo
  {author} {\bibfnamefont {P.~S.}\ \bibnamefont {Julienne}},\ }\href@noop {}
  {\bibfield  {journal} {\bibinfo  {journal} {Rev. Mod. Phys.}\ }\textbf
  {\bibinfo {volume} {71}},\ \bibinfo {pages} {1} (\bibinfo {year}
  {1999})}\BibitemShut {NoStop}%
\bibitem [{\citenamefont {Zipkes}\ \emph
  {et~al.}(2010{\natexlab{b}})\citenamefont {Zipkes}, \citenamefont {Palzer},
  \citenamefont {Ratschbacher}, \citenamefont {Sias},\ and\ \citenamefont
  {K\"{o}hl}}]{zipkes10b}%
  \BibitemOpen
  \bibfield  {author} {\bibinfo {author} {\bibfnamefont {C.}~\bibnamefont
  {Zipkes}}, \bibinfo {author} {\bibfnamefont {S.}~\bibnamefont {Palzer}},
  \bibinfo {author} {\bibfnamefont {L.}~\bibnamefont {Ratschbacher}}, \bibinfo
  {author} {\bibfnamefont {C.}~\bibnamefont {Sias}}, \ and\ \bibinfo {author}
  {\bibfnamefont {M.}~\bibnamefont {K\"{o}hl}},\ }\href@noop {} {\bibfield
  {journal} {\bibinfo  {journal} {Phys. Rev. Lett.}\ }\textbf {\bibinfo
  {volume} {105}},\ \bibinfo {pages} {133201} (\bibinfo {year}
  {2010}{\natexlab{b}})}\BibitemShut {NoStop}%
\bibitem [{\citenamefont {Raab}\ \emph {et~al.}(1987)\citenamefont {Raab},
  \citenamefont {Prentiss}, \citenamefont {Cable}, \citenamefont {Chu},\ and\
  \citenamefont {Pritchard}}]{raab87a}%
  \BibitemOpen
  \bibfield  {author} {\bibinfo {author} {\bibfnamefont {E.~L.}\ \bibnamefont
  {Raab}}, \bibinfo {author} {\bibfnamefont {M.}~\bibnamefont {Prentiss}},
  \bibinfo {author} {\bibfnamefont {A.}~\bibnamefont {Cable}}, \bibinfo
  {author} {\bibfnamefont {S.}~\bibnamefont {Chu}}, \ and\ \bibinfo {author}
  {\bibfnamefont {D.~E.}\ \bibnamefont {Pritchard}},\ }\href@noop {} {\bibfield
   {journal} {\bibinfo  {journal} {Phys. Rev. Lett.}\ }\textbf {\bibinfo
  {volume} {59}},\ \bibinfo {pages} {2631} (\bibinfo {year}
  {1987})}\BibitemShut {NoStop}%
\bibitem [{\citenamefont {Pradhan}\ and\ \citenamefont
  {Jagatap}(2008)}]{pradhan08a}%
  \BibitemOpen
  \bibfield  {author} {\bibinfo {author} {\bibfnamefont {S.}~\bibnamefont
  {Pradhan}}\ and\ \bibinfo {author} {\bibfnamefont {B.~N.}\ \bibnamefont
  {Jagatap}},\ }\href@noop {} {\bibfield  {journal} {\bibinfo  {journal} {Rev.
  Sci. Instrum.}\ }\textbf {\bibinfo {volume} {79}},\ \bibinfo {pages} {013101}
  (\bibinfo {year} {2008})}\BibitemShut {NoStop}%
\bibitem [{\citenamefont {Hall}\ \emph {et~al.}()\citenamefont {Hall},
  \citenamefont {Aymar}, \citenamefont {Bouloufa}, \citenamefont {Dulieu},\
  and\ \citenamefont {Willitsch}}]{hall11b}%
  \BibitemOpen
  \bibfield  {author} {\bibinfo {author} {\bibfnamefont {F.~H.~J.}\
  \bibnamefont {Hall}}, \bibinfo {author} {\bibfnamefont {M.}~\bibnamefont
  {Aymar}}, \bibinfo {author} {\bibfnamefont {N.}~\bibnamefont {Bouloufa}},
  \bibinfo {author} {\bibfnamefont {O.}~\bibnamefont {Dulieu}}, \ and\ \bibinfo
  {author} {\bibfnamefont {S.}~\bibnamefont {Willitsch}},\ }\href@noop {}
  {\bibinfo  {journal} {in preparation}\ }\BibitemShut {NoStop}%
\bibitem [{\citenamefont {Tacconi}\ \emph {et~al.}()\citenamefont {Tacconi},
  \citenamefont {Gianturco},\ and\ \citenamefont {Belyaev}}]{tacconi11a}%
  \BibitemOpen
\bibfield  {journal} {  }\bibfield  {author} {\bibinfo {author} {\bibfnamefont
  {M.}~\bibnamefont {Tacconi}}, \bibinfo {author} {\bibfnamefont {F.~A.}\
  \bibnamefont {Gianturco}}, \ and\ \bibinfo {author} {\bibfnamefont {A.~K.}\
  \bibnamefont {Belyaev}},\ }\href@noop {} {\bibinfo  {journal} {Phys. Chem.
  Chem. Phys.}\ ,\ \bibinfo {pages} {DOI: 10.1039/c1cp20916g}}\BibitemShut
  {NoStop}%
\bibitem [{\citenamefont {Roth}\ and\ \citenamefont
  {Schiller}(2009)}]{roth09a}%
  \BibitemOpen
\bibfield  {journal} {  }\bibfield  {author} {\bibinfo {author} {\bibfnamefont
  {B.}~\bibnamefont {Roth}}\ and\ \bibinfo {author} {\bibfnamefont
  {S.}~\bibnamefont {Schiller}},\ }in\ \href@noop {} {\emph {\bibinfo
  {booktitle} {Cold Molecules}}},\ \bibinfo {editor} {edited by\ \bibinfo
  {editor} {\bibfnamefont {R.~V.}\ \bibnamefont {Krems}}, \bibinfo {editor}
  {\bibfnamefont {W.~C.}\ \bibnamefont {Stwalley}}, \ and\ \bibinfo {editor}
  {\bibfnamefont {B.}~\bibnamefont {Friedrich}}}\ (\bibinfo  {publisher} {CRC
  Press},\ \bibinfo {year} {2009})\ p.\ \bibinfo {pages} {651}\BibitemShut
  {NoStop}%
\bibitem [{\citenamefont {Gu\'{e}rout}\ and\ \citenamefont
  {Aymar}(2010)}]{guerout10a}%
  \BibitemOpen
  \bibfield  {author} {\bibinfo {author} {\bibfnamefont {R.}~\bibnamefont
  {Gu\'{e}rout}}\ and\ \bibinfo {author} {\bibfnamefont {M.}~\bibnamefont
  {Aymar}},\ }\href@noop {} {\bibfield  {journal} {\bibinfo  {journal} {Phys.
  Rev. A}\ }\textbf {\bibinfo {volume} {82}},\ \bibinfo {pages} {042508}
  (\bibinfo {year} {2010})}\BibitemShut {NoStop}%
\bibitem [{\citenamefont {Gu\'{e}rout}\ \emph {et~al.}(2011)\citenamefont
  {Gu\'{e}rout}, \citenamefont {Aymar},\ and\ \citenamefont
  {Dulieu}}]{guerout11a}%
  \BibitemOpen
  \bibfield  {author} {\bibinfo {author} {\bibfnamefont {R.}~\bibnamefont
  {Gu\'{e}rout}}, \bibinfo {author} {\bibfnamefont {M.}~\bibnamefont {Aymar}},
  \ and\ \bibinfo {author} {\bibfnamefont {O.}~\bibnamefont {Dulieu}},\
  }\href@noop {} {\bibfield  {journal} {\bibinfo  {journal} {J. Chem. Phys.}\
  }\textbf {\bibinfo {volume} {135}},\ \bibinfo {pages} {064305} (\bibinfo
  {year} {2011})}\BibitemShut {NoStop}%
\bibitem [{\citenamefont {Bouissou}\ \emph {et~al.}(2010)\citenamefont
  {Bouissou}, \citenamefont {Durand}, \citenamefont {Heitz},\ and\
  \citenamefont {Spegelman}}]{bouissou10a}%
  \BibitemOpen
  \bibfield  {author} {\bibinfo {author} {\bibfnamefont {T.}~\bibnamefont
  {Bouissou}}, \bibinfo {author} {\bibfnamefont {G.}~\bibnamefont {Durand}},
  \bibinfo {author} {\bibfnamefont {M.-C.}\ \bibnamefont {Heitz}}, \ and\
  \bibinfo {author} {\bibfnamefont {F.}~\bibnamefont {Spegelman}},\ }\href@noop
  {} {\bibfield  {journal} {\bibinfo  {journal} {J. Chem. Phys.}\ }\textbf
  {\bibinfo {volume} {133}},\ \bibinfo {pages} {164317} (\bibinfo {year}
  {2010})}\BibitemShut {NoStop}%
\end{thebibliography}%

\end{document}